\documentclass[runningheads]{llncs}

\usepackage{amsmath}

\usepackage[T1]{fontenc}
\usepackage{booktabs}
\usepackage{adjustbox}
\usepackage{cite}
\usepackage{subcaption}
\usepackage{algorithm}
\usepackage{algpseudocode}
\usepackage{amsfonts}
\usepackage[utf8]{inputenc}
\usepackage{graphicx}
\usepackage{orcidlink}

\begin{document}
\title{A Study of Effectiveness of Brand Domain Identification Features for Phishing Detection in 2025}
%
%
\author{Rina Mishra\inst{1}\orcidID{0000-0003-2661-5858} \and
Gaurav Varshney\inst{1}\orcidID{0009-0005-5363-3519}}
\authorrunning{R. Mishra et al.}
%
\institute{Indian Institute of Technology, Jammu J\&K 181221, India \inst{1}
\email{rina.mishra@iitjammu.ac.in,gaurav.varshney@iitjammu.ac.in}}




%
\maketitle              
\begin{abstract}

Phishing websites continue to pose a significant security challenge, making the development of robust detection mechanisms essential. Brand Domain Identification (BDI) serves as a crucial step in many phishing detection approaches. This study systematically evaluates the effectiveness of features employed over the past decade for BDI, focusing on their weighted importance in phishing detection as of 2025. The primary objective is to determine whether the identified brand domain matches the claimed domain, utilizing popular features for phishing detection. Our motivation stems from the observation that certain features remain consistent across legitimate websites, while showing notable deviations in phishing websites targeting brand domains. This phenomenon is quantified and experimentally presented in this paper. To validate feature importance and evaluate performance, we conducted two experiments on a dataset comprising 4,667 legitimate sites and 4,561 phishing sites. In Experiment 1, we used the Weka tool to identify optimized and important feature sets out of 5: \texttt{CN Information(CN), Logo Domain(LD),Form Action Domain(FAD),Most Common Link in Domain(MCLD) and Cookie Domain)} through its 4-Attribute Ranking Evaluator. The results revealed that none of the features were redundant, and Random Forest emerged as the best classifier, achieving an impressive accuracy of 99.7\% with an average response time of 0.08 seconds. In Experiment 2, we trained five machine learning models—including \textit{Random Forest}, \textit{Decision Tree}, \textit{Support Vector Machine}, \textit{Multi-layer Perceptron}, and \textit{XGBoost}—to assess the performance of individual BDI features and their combinations. The results demonstrated an accuracy of \textit{99.8\%}, achieved with feature combinations of only \textit{three features}: (\texttt{Most Common Link Domain, Logo Domain, Form Action}) and (\texttt{Most Common Link Domain,CN Info,Logo Domain}) using \texttt{{Random Forest}} as the best classifier.  This study underscores the importance of leveraging key domain features for efficient phishing detection and paves the way for the development of real-time, scalable detection systems. The results offer a promising foundation for future research on enhancing detection accuracy with minimal computational overhead.
\end{abstract}
%
%
%

\section{Introduction} 
Phishing attacks is a significant and evolving threat in Cybersecurity, that use deceptive techniques to exploit user trust to steal sensitive information like Credentials, Banking details, OTP etc \cite{24}. According to the APWG's Phishing Trends Report of Q3 2024, around 80\% of phishing campaigns are primarily focused on credential theft, with cloud-based services like Microsoft 365 and Google Workspace being common targets \cite{13}. The APWG Trend Report 2024 further highlights the scale of this threat, discovering over 2.7 million unique phishing websites in just three quarters of 2024, with social media platforms emerging as a prime target \cite{11}. Additionally, TrendMicro reports a staggering 58\% increase in phishing attacks in 2023, resulting in an estimated financial loss of \$3.5 billion USD by 2024 \cite{12}. Industries such as healthcare saw a 45\% surge in phishing attacks, and government impersonation scams have increased by 35\%, showcasing the growing diversity in cybercriminals’ targets \cite{13}. Notably, credential phishing has skyrocketed by 217\% from October 2023 to March 2024 \cite{14}. These alarming statistics underscore the inadequacy of current anti-phishing measures and emphasize on an urgent need for the development of robust, efficient, and up-to-date phishing detection methods to counter the evolving tactics employed by cybercriminals \cite{15}. \\ 
Existing phishing detection techniques, such as Blacklist-based \cite{22}, Content-based \cite{23,24}, Heuristic-based \cite{20}, and Visual Similarity-based methods \cite{30}, provide valuable solutions but come with inherent limitations. Blacklist-based methods like Google Safe Browsing are efficient for identifying known threats but are ineffective against zero-day attacks or newly emerging phishing URLs. Content-based and URL feature based techniques, which extract distinguishing features from webpage content or URL structures and often utilize machine learning for classification \cite{23,25}, can face challenges like inefficiencies due to large number of features, dependency on third-party services, and potential privacy issues. Similarly, Visual similarity-based detection, are promising, but need frequent updates to handle new attacks, which increases computational overhead \cite{10}. To address these limitations, the concept of Brand Domain Identification (BDI) emerges as a compelling alternative. By systematically identifying tightly bound domain features (TBDF) that are intrinsic to legitimate websites, BDI offers a focused and efficient approach to phishing detection. A detailed systematic literature survey of these existing techniques, their limitations, and the potential of TBDF-based BDI is comprehensively discussed in section \ref{section 2}. \\
This study analyzes existing antiphishing schemes and identifies their challenges. Key challenges include an over-reliance on intricate feature engineering and extensive datasets, which impose significant computational and storage demands. Many methods are heavily dependent on third-party resources, such as Whois lookups, search engine APIs, and DNS servers, which not only introduce latency but also expose the detection pipeline to potential vulnerabilities. Additionally, the high computational complexity of these models often stems from analyzing an overwhelming number of features, resulting in inefficiencies and hindering real-time applicability. Despite the critical role of feature selection, insufficient attention has been given to understanding the relative importance of individual features, which could provide deeper insights into the fundamental mechanics of phishing detection. \\
This study aims to address the existing gap by systematically identifying and evaluating tightly bound domain features (TBDF) based on Brand Domain Identification (BDI), which are consistently associated with legitimate websites. This relationship, however, is disrupted when analyzing phishing sites targeting brand domains. The decision to focus on five critical TBDF features, including CN Info and Cookie Domain, is made to minimize reliance on third-party systems such as WHOIS, thereby enhancing the system’s independence and reliability. Phishing websites often include malicious or unrelated domains in their links, whereas legitimate websites consistently use domains associated with their brand. The Most Common Link Domain serves as a strong indicator of a website’s trustworthiness. Additionally, the domain hosting the logo is typically a unique and easily identifiable feature of a legitimate brand. Phishing sites, on the other hand, often use logos hosted on third-party or suspicious domains, making this feature an effective distinguishing factor. Furthermore, the Form Action Domain indicates where user-submitted data is sent. Phishing websites frequently use external or unrelated domains for form submissions, while legitimate websites rely on their own secure domains. By restricting the analysis to these five critical features, the study ensures computational efficiency suitable for real-time detection, with an average response time of 0.08 seconds, making it well-suited for practical deployment.


The motivation behind this study stems from the observation that certain TBDF features are consistently associated with legitimate websites, while this relationship is disrupted when dealing with phishing sites targeting brand domains. This fact has been experimentally validated and discussed in this paper. Furthermore, the review of existing literature revealed a variety of machine learning classifiers, with some favoring Random Forest \cite{17} and others XGBoost, SVM etc \cite{16,18}, sparking interest in experimenting with the outperforming classifiers identified in previous studies. We focus on the problem of analyzing the TBDF features for phishing detection, as identified in the literature. Typically, Phishing websites often replicate the textual and visual appearance of legitimate websites; however, certain features, deeply connected to a brand's identity, are challenging for phishing sites to imitate. Our problem statement centers on identifying and evaluating the significance of these features, which align with legitimate websites but exhibit mismatches in phishing sites. By leveraging these features, we aim to uncover inherent differences and enhance the robustness of phishing detection systems.
  The careful selection of TBDF features was informed by an extensive literature survey, as detailed in Section \ref{section 2}. \\  
\noindent The major contributions of this study include:
\begin{itemize}
\item To explore and assess the significance of tightly bound domain features (TBDF) in differentiating between legitimate and phishing websites.
\item Introducing a novel TBDF feature, "cookie domain," and demonstrating its practical effectiveness in detecting phishing websites.
\item Investigating the performance of TBDF features both individually and in combination, highlighting their weighted importance in enhancing phishing detection.
\item Offering valuable insights into the prioritization of features to enhance detection accuracy while minimizing computational time, thereby paving the way for more efficient and effective phishing detection systems.
\item Designing a lightweight mechanism developed for phishing detection.
\end{itemize}


The rest of the paper is organized as follows: Section \ref{section 2}  reviews related works and examines their limitations concerning feature selection for phishing detection applications. Section \ref{section 3} introduces the proposed methodology and Dataset Preparation. Section \ref{section 4} details the experimental setup, presents the results, and discusses the key findings. Lastly, Section \ref{section 5} provides the conclusion of the paper.


\section{Related Works}
{\label{section 2}}


To conduct a comprehensive literature survey, we systematically searched for the term "Brand Domain Identification Based Phishing Detection" across multiple academic databases, including Google Scholar, DBLP, IEEE Xplore, and ACM Digital Library. Our search on the ACM Digital Library yielded 2,636 results, from which only one paper met our requirements, while IEEE Xplore produced four results, of which one was relevant. DBLP returned no results, and Google Scholar produced an overwhelming 24,500 results. Given the large volume and varying relevance of papers from Google Scholar, we adopted a two-stage elimination process: first filtering papers based on their titles and then further narrowing down by carefully reviewing abstracts for relevance. Our focus was on papers published in the last 10 years, specifically those addressing brand domain identification in phishing detection and incorporating feature selection methods with promising evaluation results. Through this rigorous approach, we identified the top 20 research papers that completely matched our criteria. From these papers, we extracted TBDF features demonstrating strong associations with claimed brands in phishing detection scenarios, prioritizing those with robust evaluation metrics and significant contributions to the field.  \\
Tan et al.\cite{32}  proposed a method for phishing webpage detection by extracting identity keywords and identifying target domain names. The approach involved analyzing plain text retrieved from various HTML tags, such as meta tags, title tags, body tags, and the alt attribute of all tags, along with src and href attributes. The system achieved an accuracy of 96.1\% on a dataset of 10,000 entries. Shirazi et al. \cite{33} presented a phishing detection method that leverages various domain name-based features. These features include domain length, URL length, link ratio (the ratio of the number of hyperlinks pointing to the same domain compared to the total number of hyperlinks on the webpage), frequency of domain name occurrence, HTTPS presence, non-alphabetical characters in the domain name, domain names with copyright logos, and page title/domain name matching. The approach demonstrated an impressive accuracy range of 97-99\%, using classifiers such as SVM, SVM Gaussian, Gaussian Naive Bayes, KNN, Decision Trees (DT), Gradient Boosting (GB), and Majority Voting. The method was evaluated on a dataset containing 4,018 instances. Demidova et al. \cite{34}  introduced a hybrid approach for phishing detection that combines multiple URL-based features. These features include the number of links in <link>, <script>, <img>, and <a> tags (which can be related to brand assets, keywords, and non-brand content), the number of inputs and forms, the number of buttons on a webpage, the form methods used, and the presence of the original brand logo. This methodology achieved a remarkable 99\% accuracy, employing classifiers such as Logistic Regression (LR), XGBoost, and Random Forest. The system was tested on a large dataset consisting of more than 62,000 samples. Wang et al. \cite{35} proposed an automated phishing detection approach that utilizes Google Image Search-based logo search for identifying phishing websites. The method demonstrated an accuracy of 94.5\%, using Large Language Models (LLM) for the detection task.  Chen et al. \cite{36} proposed a method for phishing target identification by integrating multiple features such as URL features, host features (including valid days and CNAME), web resource features (links, scripts, images, and forms), login and sign-in word counts, and optical character recognition (OCR) for detecting brand logos. The method achieved an accuracy of 91.1\% and employed the PTI-NN (Phishing Target Identification Neural Network) model for classification. The approach was evaluated on a dataset consisting of 3,500 instances. Bozkir et al.  \cite{37} introduced LogoSENSE, a logo-based detection method that uses Histogram of Oriented Gradients (HOG) for identifying brand logos on phishing webpages and emails. This approach achieved an accuracy of 93.5\% and employed the SIFT (Scale-Invariant Feature Transform) technique, utilizing deep learning-based object detection for feature extraction. The model was tested on a dataset consisting of 5,039 samples. 
Li et al. \cite{38} explored the use of domain-related features for phishing detection, focusing on newly registered domains, which are often used by phishing websites. The approach achieved an accuracy of 98\% and utilized a Naive Bayes (NB) classifier for detecting phishing sites. The method was evaluated on a dataset containing 8,725 samples. Panda et al. \cite{39} proposed a novel logo identification technique aimed at detecting phishing in cyber-physical systems. The method utilizes logo-based detection and employs various classifiers, including Decision Trees (DT), Support Vector Machines (SVM), K-Nearest Neighbors (KNN), Gaussian Naive Bayes (GNB), and Extremely Randomized Forests (ERF). The technique achieved an accuracy of 87\% and was evaluated on a dataset containing 538 instances. Tan et al. \cite{40} introduced a phishing detection approach leveraging WHOIS lookups and URL-based brand features. This method assigns weights to brand names extracted from URLs to identify potential phishing attempts. The study achieved an accuracy of 98.2\% and was tested on a small dataset of 218 samples. Lee et al. \cite{41} proposed a phishing detection approach using multimodal large language models (LLMs) such as Gemini, GPT, and Claude. The method analyzes multiple features, including the title, meta description, favicon paths, and text corresponding to logo images. The approach achieved an accuracy of 90\% on a dataset comprising 4,480 samples.  Tan et al. \cite{42}  proposed a hybrid approach to phishing detection that combines visual features, such as logos, with textual features like brand-related keywords. This method aims to enhance detection accuracy by leveraging the synergy between visual and textual identities. The approach achieved an accuracy of 98.6\% on a dataset consisting of 1,250 samples. Xiang et al.  \cite{43} presented a hybrid method for phishing detection that incorporates multiple features, including title, copyright information, named entity recognition (NER), login form detection, and keyword-based retrieval. The approach achieved an accuracy of 90.06\% on a dataset of 11,449 samples. Ramesh et al. \cite{44}  proposed a phishing detection method based on analyzing the relationship between phishing webpages and their target domains. The approach focuses on detecting login pages and leveraging URL-based features to identify phishing attempts. The method achieved an impressive accuracy of 99\% on a dataset of 3,675 samples. Sahingoz et al.  \cite{45} introduced a phishing detection framework leveraging URL-based features to classify websites as phishing or legitimate. The study employed various machine learning classifiers, including Decision Trees (DT), AdaBoost, Kstar, k-Nearest Neighbors (kNN), Random Forest (RF), Sequential Minimal Optimization (SMO), and Naïve Bayes (NB). The method achieved an accuracy of 97.98\% on a substantial dataset of 73,575 samples. \\
 Building upon insights gathered from an extensive literature survey, we identified several critical gaps that warrant further investigation in the domain of phishing detection which are as follows:

 \begin{itemize}
    \item  There is limited systematic evaluation of the effectiveness of Brand Domain Identification (BDI)-based features over time.
    \item Many schemes are computationally expensive due to the inclusion of a large number of features, and may also require the extraction of complex features.
    \item Schemes that rely on third-party features are dependent on their accuracy and availability.
    \item Previous work lacks an emphasis on understanding the weighted importance of features in phishing detection.
 \end{itemize}

Addressing these gaps is imperative for advancing phishing detection methodologies and ensuring robust, efficient, and scalable solutions.

\section{Proposed Methodology}
\label{section 3}

\begin{figure}[!htbp]
    \centering
    \includegraphics[width=1\linewidth]{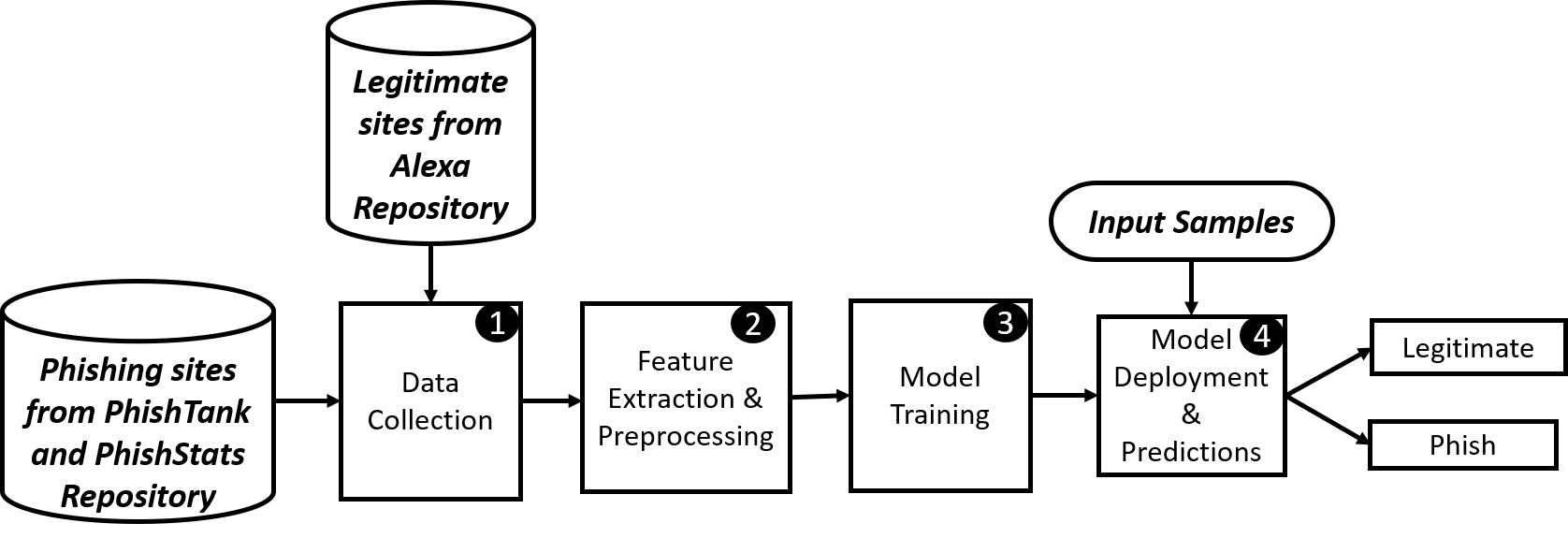}
    \caption{Proposed Methodology}
    \label{fig:1}
\end{figure}

\begin{figure}[htbp]
    \centering
    \begin{subfigure}[t]{0.55\linewidth}
        \centering
        \includegraphics[width=\linewidth]{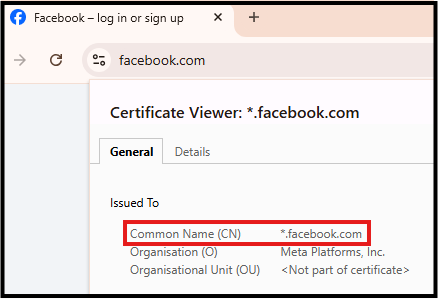}
        \caption{CN Info}
        \label{fig:CN_}
    \end{subfigure}
    \hfill
    \begin{subfigure}[t]{0.85\linewidth}
        \centering
        \includegraphics[width=\linewidth]{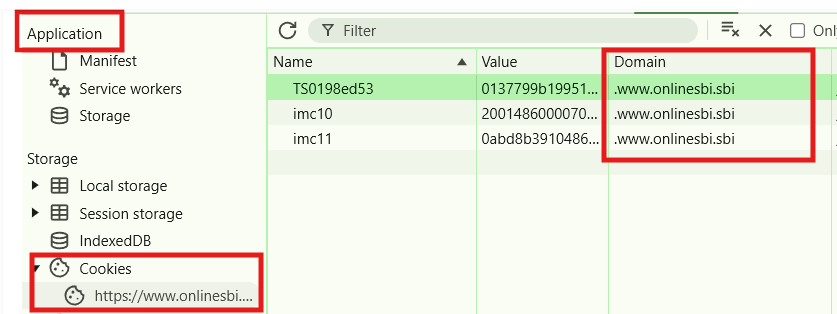}
        \caption{Cookie Domain}
        \label{fig:CD_}
    \end{subfigure}
    
    \vspace{0.5cm}
    \begin{subfigure}[t]{0.99\linewidth}
        \centering
        \includegraphics[width=\linewidth]{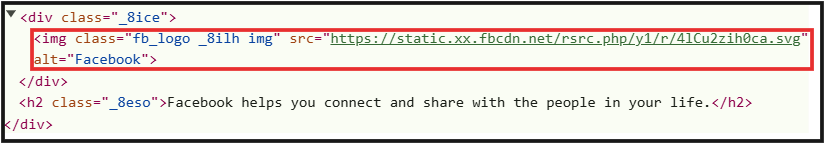}
        \caption{Logo Domain}
        \label{fig:LD_}
    \end{subfigure}
    \hfill
    \begin{subfigure}[t]{0.85\linewidth}
        \centering
        \includegraphics[width=\linewidth]{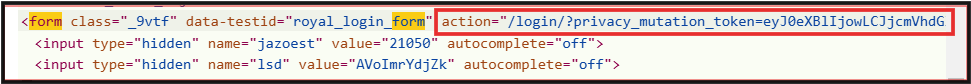}
        \caption{Form Action Domain}
        \label{fig:FAD_}
    \end{subfigure}
    
    \vspace{0.5cm}
    \begin{subfigure}[t]{0.99\linewidth}
        \centering
        \includegraphics[width=\linewidth]{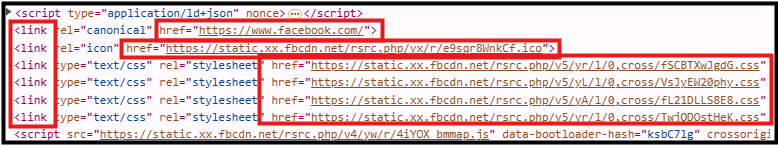}
        \caption{Most Common Link in Domain}
        \label{fig:MCLD_}
    \end{subfigure}
    
    \caption{Visualization of Various Features Related to Domain Information}
    \label{fig:Domain_Features}
\end{figure}

To address the research gaps identified in section \ref{section 2}, we are proposing a systematic study on Brand Domain Identification (BDI) based TBDF features utilized over the past decade, emphasizing their relevance to phishing detection. This section describes proposed methodology, Dataset preparation and Model deployment details as shown in Figure \ref{fig:1}. \\
The first step of Figure \ref{fig:1} involved \textit{Data Collection(1)}, during which legitimate and phishing data were gathered from different sources discussed in detail in Subsection \ref{DS Preparation}. This was followed by \textit{Feature Extraction \& Preprocessing(2a)} phase, in which TBDF features like \texttt{CN Info, CD,LD,MCLD and FAD} gets extracted. For example consider Figure \ref{fig:Domain_Features}  and Table \ref{table:feature_extraction} if a user visits a website www.facebook.com. For this site the domain extracted will be facebook.com, which is also identified as the root domain in this case. This information is retrieved using the tldextract Python library \cite{49} and stored into domain and root domain variable of Table \ref{table:feature_extraction} respectively. The Common Name column captures the Common Name (CN) extracted from the SSL/TLS certificate issued for the domain which can be visualized from Figure \ref{fig:CN_}, and gets stored in CN Info variable of Table \ref{table:feature_extraction}. For facebook.com, the CN value is  \textit{*.facebook.com}. To ensure consistency and accuracy in matching, extraneous symbols such as * or www. are removed, resulting in a cleaned CN value of facebook.com.
The Cookie Domain (CD) feature is derived from the domain that appears most frequently within the cookie record represented in Figure \ref{fig:CD_} Cookie Domain. For facebook.com, its value is \textit{.facebook.com}. The leading dot (.) is removed to standardize the feature, yielding a final CD value of facebook.com which is stored under Cookie Domain variable of Table \ref{table:feature_extraction}. Similarly, the Most Common Link Domain represents the predominant domain found within the <a> tags of the website’s HTML content highlighted in Figure \ref{fig:MCLD_}, which in this case is \textit{facebook.com}. The Logo Domain feature represented in Figure \ref{fig:LD_} captures the domain, most frequently linked to the website’s logo, which, in this example, belongs to facebook.com and stored under Logo Domain variable of Table \ref{table:feature_extraction}. Finally, the Form Action Domain identifies the primary domain associated with form submission endpoints on the site highlighted by Figure \ref{fig:FAD_}.

After extracting these features, \textit{preprocessing(2b)} is required to prepare them for training machine learning models, as described in Table \ref{tab:traditional_features}. For instance, for www.facebook.com, the root domain: facebook.com was matched against the CN Info extracted from the SSL certificate. If this domain matched the CN Info (i.e., both were facebook.com), it was assigned a value of 1. Conversely, if a mismatch occurred, such as with the domain github-facebook.com, it was assigned a value of -1. If the feature was absent, it was assigned a value of 0.


The preprocessing phase, which included eliminating missing entries and converting the data into numeric form (as shown in Table \ref{tab:traditional_features}), culminated in the creation of a dataset consisting of \texttt{9228 websites and five features}. This structured approach provided a solid foundation for evaluating the effectiveness of the selected features in phishing detection.

The next step involved \textit{Model Training(3)}. We performed the training by dividing the dataset into an 80:20 ratio for training and testing, respectively. This means that 80\% of the data was used to train the models, while the remaining 20\% was reserved for \textit{prediction(4)}. We selected these classifiers due to their proven effectiveness and versatility, as demonstrated in prior literature \cite{46,15,47,27,48}. Collectively, these classifiers are well-supported by the literature for their applicability and reliability in addressing a wide range of machine learning challenges, making them ideal for our study.

\subsection{Data Collection and Dataset Preparation}
\label{DS Preparation}
To create a dataset comprising \textit{4667 legitimate websites} and \textit{4561 phishing websites}, Table \ref{table:feature_extraction} was referenced for feature extraction, and Table \ref{tab:traditional_features} was used to convert these features into a format compatible with machine learning. If the feature matched the domain or root domain of the webpage, it got a value of 1. If it didn’t match, it got -1. If the feature was missing, it got 0.

\noindent \textbf{Legitimate Websites}: A dataset of \textit{4667 legitimate websites} was curated from the \texttt{Alexa Top 1 Million Dataset} \cite{26}. Only websites that allowed crawling, did not block feature extraction, and for which at least three feature values could be fetched were included, ensuring a diverse and representative sample of legitimate domains. Since the Alexa dataset has been widely used by previous researchers as a source for normal datasets, we incorporated it into our experiment to maintain transparency and consistency. These entries were labeled as 'F' in the Dataset. \\
\noindent \textbf{Phishing Websites}: \texttt{PhishTank} \cite{19} and \texttt{PhishStats} \cite{29} are two widely recognized real-time datasets of phishing sites actively operating on the internet. Suspected phishing sites are contributed by a diverse range of users and third parties worldwide, with their validity determined through votes cast by other users or experts. For our study, data was collected from PhishTank  and PhishStats  between November 20, 2024, and January 10, 2025. Only sites that were active and accessible during the data collection period, and for which at least three feature values could be fetched, were included in the dataset. These entries were labeled as 'T' in the Dataset. \\

\begin{table}
\caption{TBDF-Based Feature Extraction for Dataset Preparation}
\label{table:feature_extraction}
\centering
\begin{tabular}{|p{3.5cm}|p{3cm}|p{4cm}|}
\hline
\textbf{Feature}                & \textbf{Description}                                                    & \textbf{Methodology}                                                                                                     \\ \hline
\textbf{Domain from URL}         & Extracted domain, including subdomain, domain name, and suffix.         & Used \texttt{tldextract} \cite{49} to parse and extract the full domain including subdomains.  Example:\texttt{sub.example.com}                                         \\ \hline
\textbf{Root Domain from URL}    & Extracted root domain (domain + suffix).                                & Used \texttt{tldextract}\cite{49} to isolate the root domain from the full domain. Example: \texttt{example.com}     \\ \hline
\textbf{CN Info}      & Common Name (CN) field from the SSL certificate of the domain.          & Used \texttt{openssl}\cite{50} to fetch the SSL certificate and parsed the CN field from the certificate's subject information. Example: \texttt{example.com}     \\ \hline
\textbf{Cookie Domain}           & Extracted domain field from HTTP cookies returned by the server.        & Parsed cookies using \texttt{requests}\cite{51} library and cleaned the domain by removing prefixes such as \texttt{www.} or \texttt{*.}. Example:\texttt{example.com}     \\ \hline
\textbf{Most Common Link in Domain} & Most frequently occurring domain among hyperlinks (\texttt{<a>} tags). & Used \texttt{BeautifulSoup}\cite{52} to parse webpage content, extracted all hyperlink domains, and identified the most common one. Example: \texttt{example.com}     \\ \hline
\textbf{Form Action Domain}      & Domain extracted from the \texttt{action} attribute of HTML \texttt{<form>} tags. & Parsed \texttt{<form>} tags with \texttt{action} attributes using \texttt{BeautifulSoup}\cite{52} and selected the longest domain among them. Example: \texttt{form.example.com} \\ \hline
\textbf{Logo Domain}             & Domain of the most common logo images on the webpage.                  & Extracted \texttt{<img>} tags with \texttt{src} containing keywords like "logo," resolved absolute paths, and identified the most frequently occurring domain. Example: \texttt{asts.example.com} \\ \hline
\end{tabular}
\end{table}



\begin{table}[!htbp]
\centering
\caption{Traditional Feature-Based Description and Value Range}
\resizebox{\textwidth}{!}{%
\begin{tabular}{p{6cm}p{1.5cm}p{6cm}}
\hline
\textbf{Feature} & \textbf{Value Range} & \textbf{Description} \\ \hline
Form Action Detection(FAD) & 0, 1, -1 & 0: Feature absent, 1: Identified Domain matches Domain or root domain, -1: Identified Domain mismatches either of them. \\ \hline
Logo Domain(LD) & 0, 1, -1 & 0: Feature absent, 1: Identified Domain matches Domain or root domain, -1: Identified Domain mismatches either of them. \\ \hline
CN Information(CN) & 0, 1, -1 & 0: Feature absent, 1: Identified Domain matches Domain or root domain, -1: Identified Domain mismatches either of them. \\ \hline
Most Common Link in Domain (MCLD) & 0, 1, -1 & 0: Feature absent, 1: Identified Domain matches Domain or root domain, -1: Identified Domain mismatches either of them. \\ \hline
Cookie Domain(CD) & 0, 1, -1 & 0: Feature absent, 1: Identified Domain matches Domain or root domain, -1: Identified Domain mismatches either of them. \\ \hline
\end{tabular}%
}
\label{tab:traditional_features}
\end{table}
\newpage
\subsection{Model Training}
We deployed 5 Machine Learning Models for Experiment 2 for phishing detection. Although the best model identified in Experiment 1 was Decision Tree so we are using Decision Tree for model training. Additionally we are training other popular models in the literature to get the best predictions. The following models are used in the experiments:\\
\textbf{XGBoost} : XGBoost is an optimized gradient boosting library designed for efficiency, flexibility, and scalability. It enhances performance by supporting distributed systems and does not rely on linear features or feature interactions. Additionally, it incorporates a regularization framework to reduce overfitting, making it highly effective for classification and regression tasks \cite{46}.  \\ 
\textbf{MLP (Multi-Layer Perceptron)} : A Multilayer Perceptron (MLP) is a widely used neural network comprising an input layer, hidden layer(s), and an output layer. The input layer receives signals, the hidden layer processes them, and the output layer makes predictions. MLPs are commonly applied to supervised learning tasks to learn input-output relationships\cite{15}.   \\ 
\textbf{Random Forest: }
Random Forest (RF) is an ensemble method that uses bagging to build multiple decision trees and predicts the output by majority voting. Randomness in feature selection and node splitting reduces overfitting and enhances diversity, making RF efficient and fast. The final prediction is the average of individual tree outputs \cite{47}.  \\ 
\textbf{Decision Tree:}
This model represents data using a rooted tree structure that consists of nodes, edges, and leaves. Each node is labeled with features, edges are marked with corresponding feature values, and the leaves are tagged with class labels. When classifying an instance with an unknown class, the process begins at the root node, where the instance is directed through the tree based on its feature values. At each node, a test is performed on the feature values, determining the path along the edges until reaching a leaf node. The classification outcome is determined by the label of the leaf node where the process concludes\cite{27}.  \\
\textbf{SVM (Support Vector Machine):}  (SVM) algorithm is designed to identify a hyperplane in an N-dimensional space, where N represents the total number of features. The dimensions of the hyperplane are determined by the number of features. SVM's primary objective is to maximize the margin between the hyperplane and the data points, ensuring clear separation between classes. By determining the optimal decision boundary, SVM enables accurate classification of data points into their respective categories, facilitating the placement of new data points in the appropriate class with ease\cite{48}. 
\section{Experiments and Results}
\label{section 4}
This section describes experimental setup, evaluation matrix utilized to test the results, two experiments conducted in this study, along with the results obtained. Experiment 1 is performed using Weka tool for feature selection and model selection and Experiment 2 is performed to assess the importance of TBDF features using five machine learning models and to evaluate the performance of each feature and their combinations to identify their importance in phishing detection.
\subsection{Experimental Setup}
The experiments were conducted on an Ubuntu 20.04 64-bit virtual machine equipped with 75 GB of storage space, 4 CPU cores, and 8 GB of RAM. Python was used as the primary programming language, and the development was performed in Visual Studio Code. This setup was designed to provide an efficient and controlled environment for feature evaluation, ensuring reliable results across both categories of websites.
\subsection{Evaluation Matrix}
During classification, phishing websites correctly identified are labeled as True Positive (TP), while those mistakenly classified as legitimate are False Negatives (FN). Legitimate websites correctly identified are True Negatives (TN), and those erroneously classified as phishing are False Positives (FP). We report standard classification metrics, including: 
\begin{equation}
\text{TPR} = \frac{TP}{TP + FN}
\end{equation}

\begin{equation}
\text{FPR} = \frac{FP}{FP + TN}
\end{equation}

\begin{equation}
\text{Precision} = \frac{TP}{TP + FP}
\end{equation}

\begin{equation}
\text{Recall} = \frac{TP}{TP + FN}
\end{equation}

\begin{equation}
\text{F-Measure} = 2 \cdot \frac{\text{Precision} \cdot \text{Recall}}{\text{Precision} + \text{Recall}}
\end{equation}

\begin{equation}
\text{Accuracy} = \frac{TP + TN}{TP + TN + FP + FN}
\end{equation}
\subsection{Experiment 1}
The primary objective of this experiment is to perform feature selection and model selection. Feature selection is essential for identifying the most significant features while eliminating redundant ones, thereby enhancing the efficiency and accuracy of the models. Model selection focuses on determining the best-performing model for the chosen feature set, ensuring optimal predictive performance.
For this purpose, Weka 3.8.6 was employed as the tool for both feature selection and model selection. Weka, a comprehensive suite of machine learning algorithms, facilitates various data mining tasks, including data preparation, classification, regression, clustering, association rule mining, and visualization \cite{38}. Its versatility and ease of use make it an ideal choice for implementing the experiments in this study.
\subsubsection{Feature Selection:}
Feature selection is a pivotal process in machine learning that aims to identify a subset of relevant and informative features from a larger set of input attributes. The primary goal is to select the most suitable features that are unbiased, improve the accuracy and efficiency of machine learning models, and reduce dimensionality. This process involves eliminating redundant or irrelevant features, minimizing model complexity, and prioritizing the most informative attributes \cite{10}. In this study, feature selection is conducted using the filter method.
Filter methods select features based on performance metrics without depending on the specific data modeling algorithm or predictor. Once the optimal features are identified, they can be utilized by the chosen modeling algorithm \cite{15}.
We employed four popular filter methods commonly cited in the literature for feature selection: 1) CorrelationAttributeEval assesses the correlation between each attribute and the class to determine its relevance and can be used to rank attributes by their correlation scores. 2) InfoGainAttributeEval evaluates the worth of an attribute by measuring the information gain with respect to the class and ranks attributes based on their information gain values. 3) GainRatio is a variation of Information Gain that adjusts for bias towards attributes with more values. It evaluates the significance of an attribute using a normalized gain measure. 4) The Relief algorithm measures the quality of attributes based on how well their values differentiate between instances that are near to each other\cite{54}. These methods collectively ensure that the selected features are both informative and effective for building robust machine learning models.
\begin{table}[H]
\centering
\scriptsize 
\caption{Attribute Ranking Based on Different Evaluators}
\label{Tab:Attribute_Ranking}
\resizebox{\textwidth}{!}{%
\begin{tabular}{@{}p{6cm}p{1.5cm}p{1.5cm}p{1.5cm}p{1.5cm}p{1.5cm}@{}}
\toprule
\textbf{Attribute Evaluator} & \textbf{F1} & \textbf{F2} & \textbf{F3} & \textbf{F4} & \textbf{F5} \\ \midrule

\textbf{CorrelationAttributeEval}& MCLD  &LD & FAD & CN & CD \\
\textit{Rank}                           & 0.85             & \textbf{0.854}                  & 0.85 & 0.848       & 0.842      \\ \midrule
\textbf{GainRatioAttributeEval}  & MCLD &LD & FAD & CN & CD\\
\textit{Rank}                           & 0.451             & \textbf{0.453}                   & 0.448  & 0.451       & 0.44      \\ \midrule
\textbf{InfoGainAttributeEval}   & MCLD &LD & CN & FAD & LD \\
\textit{Rank}                           & \textbf{0.709}              & 0.708                   & 0.707 & 0.703       & 0.693      \\ \midrule
\textbf{ReliefFAttributeEval}      & MCLD &LD & FAD & CD & CN\\
\textit{Rank}                           & \textbf{0.301}   &0.133         & 0.219                  & 0.215 & 0.174           \\ \midrule
\end{tabular}%
}
\end{table}
The results summarized in Table \ref{Tab:Attribute_Ranking} suggest that all Attribute Evaluators are selecting all features, but they show different rankings for each individual feature. This indicates that the feature combination is optimized and that none of the features in the set are redundant. Additionally, according to the results in Table \ref{Tab:Attribute_Ranking}, the feature LD contributes the most in terms of Correlation, with the highest rank value of 0.854. Similarly, in the GainRatio method, LD is the best feature, with a rank of 0.453. In InfoGain, MCLD is the best feature, with a rank of 0.709, while in ReliefF, MCLD also emerges as the best feature, with a rank of 0.301. The differences in ranking values arise from the distinct evaluation methods used by these filters.
\subsubsection{Classifier selection:}
The problem addressed in this study is a classification task, where we aim to determine whether a given website is phishing or legitimate. Selecting the best classifier is, therefore, a critical aspect of the solution.
The experiments utilized the 10-fold cross-validation technique to evaluate the models, as it effectively reduces estimation variance. This method involved splitting the training dataset into 10 subsets. In each iteration, one subset served as the test set, while the remaining nine subsets were used for training. Each subset acted as the test set exactly once across the 10 iterations, ensuring a thorough evaluation of the entire dataset.
Based on the results presented in Table \ref{tab:8_CL}, the Random Forest classifier was identified as the best performer. Since all performance metrics yielded comparable results, the selection was ultimately made based on the classifier's minimal prediction time, making Random Forest the most efficient choice for this study.

\begin{table}[H]
\centering
\caption{Classifier considering all Performance Metrics}
\label{tab:8_CL}
\begin{adjustbox}{max width=\textwidth}
\large
\begin{tabular}{lcccccc}
\toprule
\textbf{Classifier} & \textbf{TP Rate} & \textbf{FP Rate} & \textbf{Precision} & \textbf{Recall} & \textbf{F-Measure} & \textbf{Time (s)} \\
\midrule
NaiveBayes & 0.997 & 0.003  & 0.997 & 0.997 & 0.997 & 0.14 \\
AdaBoostM1 &0.997 & 0.003  & 0.997 & 0.997 & 0.997 & 0.23  \\
J48 & 0.997 & 0.003 & 0.997 & 0.997 & 0.997 & 0.18  \\
\textbf{Random Forest} & \textbf{0.997} & \textbf{0.003} & \textbf{0.997} & \textbf{0.997} & \textbf{0.997}& \textbf{0.08} \\
\bottomrule
\end{tabular}
\end{adjustbox}
\label{tab:classifier_performance}
\end{table}

\subsection{Experiment 2}
In this experiment, five machine learning models were trained and tested using individual TBDF features as well as combinations of these features. The accuracy was determined based on the performance of the best model among the five. For example, Table \ref{tab:phishing_analysis} presents results where all five models were trained using only the CN Info TBDF feature, achieving an accuracy of 87.4\%, with the Random Forest classifier emerging as the top performer. Similarly, the accuracy for other TBDF features—CD, MCLD, FAD, and LD—was computed across all classifiers, with the highest accuracy and the corresponding model recorded in Table \ref{tab:phishing_analysis}.

Next, combinations of two TBDF features were used to train all five models, with the best-performing classifier and its corresponding accuracy documented. This process was repeated for combinations of three, four and five features, with the best-performing model and its accuracy recorded at each stage in Table \ref{tab:phishing_analysis}. The first column of the table shows the number of features used, the second column lists the feature names, the third column indicates the accuracy, and the fourth column identifies the best classifier out of the five.

Our search concluded after achieving an exceptional detection accuracy of 99.8\% with a combination of three features: Most Common Link Domain, CN Info, and Form Action, as well as CN Info, Form Action, and Logo Domain. This same accuracy was also achieved with combinations of four and five features, but the combination of three features proved to be the most optimized. Therefore, this feature combination is identified as the best based on our experiments.

\begin{figure}[H]
    \centering
    \includegraphics[width=0.7\linewidth]{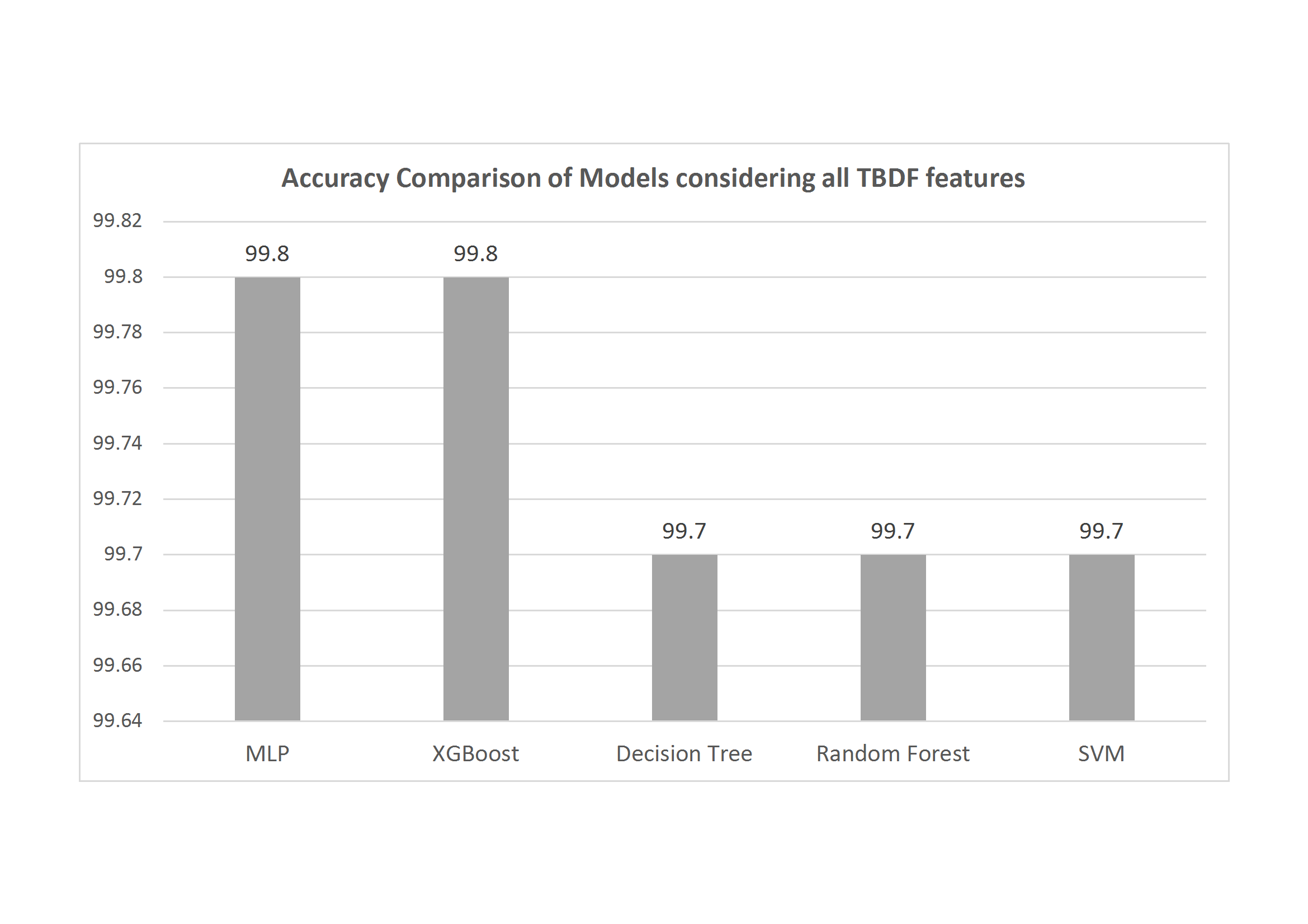}
    \caption{Accuracy Comparison of various models considering all TBDF features.}
    \label{fig:accuracy}
\end{figure}

\begin{figure}[H]
    \centering
    \includegraphics[width=0.8\linewidth]{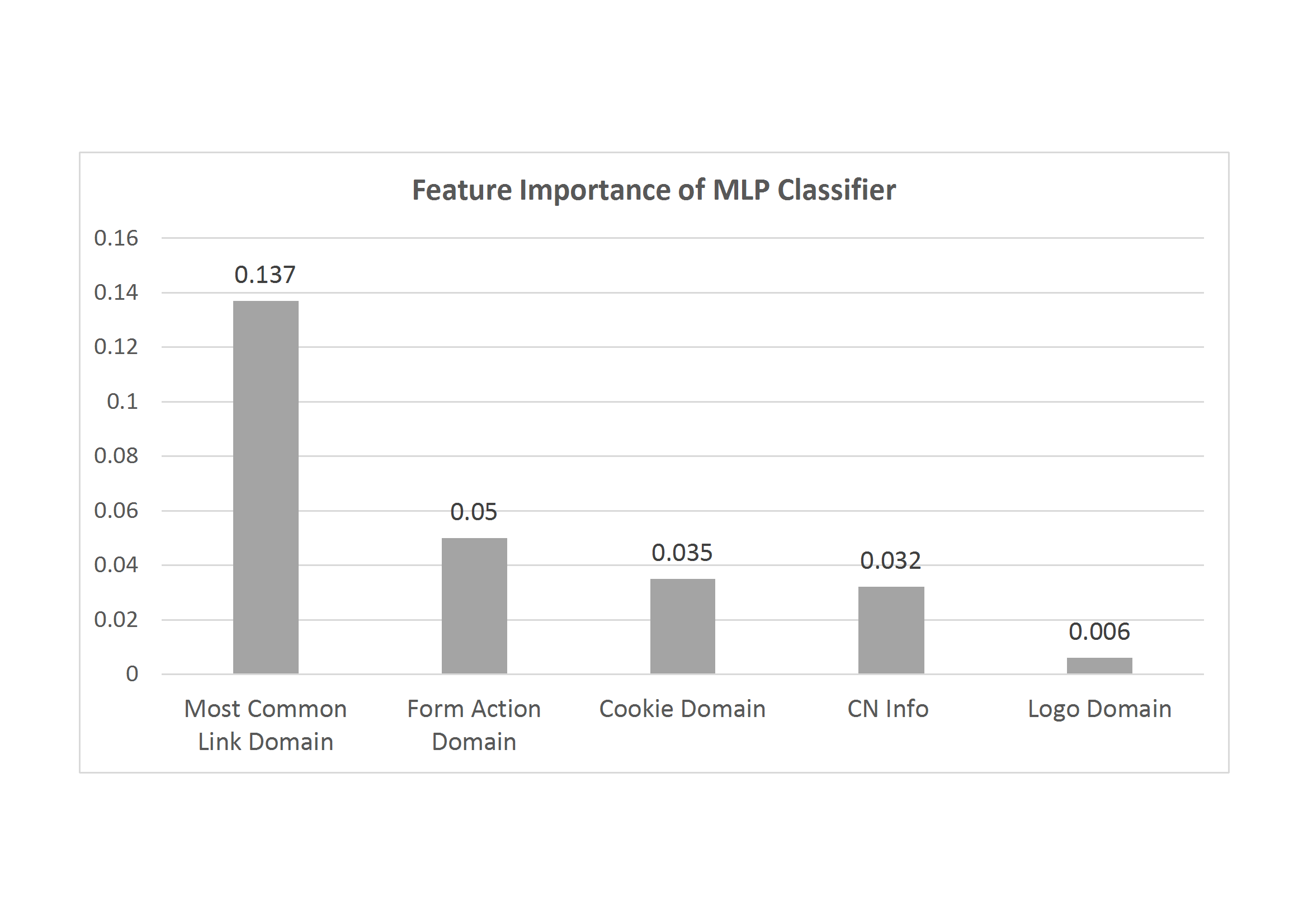}
    \caption{Feature Importance Comparison of MLP classifier}
    \label{fig:compare_all}
\end{figure}

Figure \ref{fig:accuracy} compares the accuracy of all classifiers using different TBDF feature combinations, while Figure \ref{fig:compare_all} emphasizes the importance of each TBDF feature. The results indicate that the Most Common Link Domain is the most significant feature.
\begin{table}
\centering
\scriptsize 
\caption{\label{tab:phishing_analysis}\textbf{Feature Combinations and Accuracy}}
\begin{tabular}{|c|l|c|l|}
\hline
\textbf{Comb.} & \textbf{Feature Combination} & \textbf{Acc (\%)} & \textbf{Model} \\ \hline
1 & CN Info & 87.4 & Random Forest \\ \hline
1 & Cookie Domain & 85.6 & Random Forest \\ \hline
1 & Most Common Link Domain & 87.8 & Random Forest \\ \hline
1 & Form Action Domain & 86.5 &  Random Forest \\ \hline
1 & Logo Domain & 87.6 & MLP \\ \hline
2 & CN Info, Cookie Domain & 96.5 & Random Forest \\ \hline
2 & Most Common Link, CN Info & 96.4 & Random Forest \\ \hline
2 & CN Info, Form Action & 96.7 & Random Forest \\ \hline
2 & CN Info, Logo Domain & 97.7 & SVM \\ \hline
2 & Most Common Link, Cookie & 95.8 & Random Forest \\ \hline
2 & Cookie Domain, Form Action & 96.7 & Random Forest \\ \hline
2 & Cookie, Logo Domain & 97.1 & MLP \\ \hline
2 & Most Common Link, Form Action & 97.8 & Random Forest \\ \hline
2 & Most Common Link, Logo Domain & 97.2 & Random Forest \\ \hline
2 & Form Action, Logo Domain& 97.0 & Random Forest \\ \hline
3 & Most Common Link, CN Info, Cookie & 99.1 & Random Forest \\ \hline
3 & CN Info, Cookie, Form Action & 99.1 & Random Forest \\ \hline
3 & CN Info, Cookie, Logo Domain& 99.7 & Random Forest \\ \hline
3 & Most Common Link, CN Info, Form Action & 99.1 & SVM \\ \hline
3 & Most Common Link, CN Info, Logo Domain & 99.8 & Random Forest \\ \hline
3 & CN Info, Form Action, Logo Domain & 99.8 & Random Forest \\ \hline
3 & Most Common Link, Cookie, Form Action & 99.7 & Random Forest \\ \hline
3 & Most Common Link, Cookie, Logo Domain& 99.2 & Random Forest \\ \hline
3 & Cookie, Form Action, Logo Domain& 99.4 & Random Forest \\ \hline
3 & Most Common Link, Form Action, Logo Domain& 99.7 & Random Forest\\ \hline
4 & Most Common Link, CN Info, Cookie, Form Action & 99.8 & Random Forest \\ \hline
4 & Most Common Link, CN Info, Cookie, Logo Domain& 99.8 & Random Forest \\ \hline
4 & CN Info, Cookie, Form Action, Logo Domain & 99.7 & Decision Tree \\ \hline
4 & Most Common Link, CN Info, Form Action, Logo Domain & 99.8 & SVM \\ \hline
4 & Most Common Link, Cookie, Form Action, Logo Domain& 99.6 & Random Forest \\ \hline
5 & Most Common Link, CN Info, Cookie, Form Action, Logo Domain& 99.8 & MLP\\ \hline
\end{tabular}
\end{table}
\vspace{-1.5cm}

Figure \ref{fig:accuracy}  highlights MLP and XGBoost as the top-performing models, achieving a remarkable testing accuracy of 99.8\% when all features were considered together. In contrast, the remaining four machine learning models demonstrated similar accuracies of 98.7\%, showcasing their effectiveness but slightly underperforming compared to MLP and XGBoost.


\subsection{Comparision with Related Work}

We conducted an empirical comparison of our results with existing state-of-the-art solutions, as summarized in Table \ref{tab:comparison}. The evaluation focused on key aspects such as dataset size, the number of features utilized, and the accuracy achieved. Notably, we excluded run-time metrics from the comparison, as these are system-specific and may vary significantly. The comparison was conducted with these state-of-the-art methods because the foundational ideas of our study were inspired by these works.

The insights derived from Table \ref{tab:comparison} highlight that our approach leverages a larger dataset than those used in the compared methods, employs a more compact feature set, and yet achieves superior accuracy. This validates the significance of our study, underscoring the efficacy and impact of our proposed methodology.


\begin{table}[htbp]

\centering
\caption{Comparison with State-of-the-art Approaches}
\label{tab:comparison}
\begin{tabular}{|l|c|c|c|c|l|}
\hline
\textbf{Approach}              & \textbf{Legt Size} & \textbf{Phish Size} & \textbf{Feature Set} & \textbf{Accuracy} \\ \hline
Cantina \cite{55}                  & 2100                           & 19                           & 7                       & 96.97\%                             \\ \hline
Cantina+ \cite{23}                 & 1868                           & 940                          & 15                      & 97\%                            \\ \hline
Our approach &    4667                       & 4561                 & 5                       & 99.8\%                        \\ \hline
\end{tabular}
\end{table}
\section{Conclusions and Future Work}
\label{section 5}

This study emphasizes the critical role of Brand Domain Identification (BDI) as an initial step toward phishing detection. By systematically evaluating the weighted importance of features over the past decade, our work validates their effectiveness in determining whether an identified brand domain aligns with the claimed domain. Through extensive experimentation using a dataset of 4667 legitimate and 4561 phishing websites, our study highlights the potential of optimized feature selection in improving detection accuracy. Experiment 1 leveraged the Weka tool to identify the most important features and optimize classifiers, resulting in Random Forest achieving a remarkable accuracy of 99.7\% with an average response time of 0.08 seconds. In Experiment 2, we further demonstrated that a minimal feature set of just three attributes (Most Common Link Domain, Logo Domain, Form Action) yielded an exceptional accuracy of 99.8\% using Random Forest, showcasing the efficiency and robustness of our approach. Comparison with state-of-the-art methods underscores the significance of our methodology, as we achieved superior accuracy with a smaller, optimized feature set while using a larger dataset. This validation reaffirms the relevance of our study in advancing phishing detection mechanisms. Building on the promising results of this study, future work will focus on integrating the proposed feature set and classifiers into real-time phishing detection systems, ensuring scalability and robust performance against evolving phishing tactics, including zero-day attacks. We aim to optimize computational efficiency for handling larger datasets and explore cross-domain applications such as scam and fraudulent domain detection. Additionally, incorporating explainable AI and advanced ensemble techniques will enhance transparency and further improve accuracy, contributing to more reliable and adaptive phishing detection mechanisms.

\begin{credits}

\subsubsection{\discintname} The authors have no competing interests to declare that are
relevant to the content of this article. 
\end{credits}
%
%
%
%





\end{document}